# Agile System Development Lifecycle for AI Systems: Decision Architecture

Asif Q. Gill, *School of Computer Science, Faculty of Engineering & IT, University of Technology Sydney, Ultimo, NSW 2007, Australia*

*Abstract*—**Agile system development life cycle (SDLC) focuses on typical functional and non-functional system requirements for developing traditional software systems. However, Artificial Intelligent (AI) systems are different in nature and have distinct attributes such as (1) autonomy, (2) adaptiveness, (3) content generation, (4) decision-making, (5) predictability and (6) recommendation. Agile SDLC needs to be enhanced to support the AI system development and ongoing post-deployment adaptation. The challenge is: how can agile SDLC be enhanced to support AI systems? The scope of this paper is limited to AI system enabled decision automation. Thus, this paper proposes the use of decision science to enhance the agile SDLC to support the AI system development. Decision science is the study of decision-making, which seems useful to identify, analyse and describe decisions and their architecture subject to automation via AI systems. Specifically, this paper discusses the decision architecture in detail within the overall context of agile SDLC for AI systems. The application of the proposed approach is demonstrated with the help of an example scenario of insurance claim processing. This initial work indicated the usability of a decision science to enhancing the agile SDLC for designing and implementing the AI systems for decision-automation. This work provides an initial foundation for further work in this new area of decision architecture and agile SDLC for AI systems.**

## I. INTRODUCTION

SOFTWARE SYSTEMS are core to our increasingly digitally-enabled economy and society. Agile methods provide an iterative and incremental approach to deal with the complex undertaking of software development, especially when the requirements are not fixed or subject to change. While agile methods got lot of attention from academia and industry since early 2000, however, the history of agile software development can be traced back to 1950s when iterative and incremental approach was applied to the X-15 hypersonic jet and project mercury (NASA) [1]. Over the period of five decades (1950-2000), agile software development practices evolved to address the challenges of traditional waterfall and incremental approaches. Agile Alliance formalized agile software development through the formation of Manifesto for Agile

Software Development in 2001 [2]. This manifest provides four core agile values and twelve principles to guide agile software development. It has been a while since this manifesto was proposed, however, it is still relevant to agile software development practices across the globe. A number of agile methods have been proposed and evolved over a period of time that underpin agile values and principles such as adaptive software development [3], behavior-driven development [4], crystal methodologies [5], dynamic software development [6], disciplined agile delivery [7], extreme programming [8], feature-driven development [9], lean software development [10], scrum [11], scaled agile framework [12], and test-driven development [13]. Agile software development was further integrated into operations, which led to the emergence of DevOps approach enabling the continuous integration and delivery of software in production as quickly as possible for end user consumption [14].

### A. Research Motivation and Gap

Agile methods have been intensively researched in the last two decades (e.g. [15-23]) to deal with the evolving software development requirements. While they seem to work fine for traditional functional and non-functional software system requirements, however, artificial intelligence (AI) systems are different in nature and demand new or rethinking [24]. AI systems heavily rely on data processing for intelligence and can behave in an unpredictable manner post deployment [24, 54]. Similar to agile, the history of AI systems can also be traced back to 1950s [25]. There are several definitions of AI systems. This paper adopted the generally accepted, yet comprehensive definition of AI systems from OECD (Organisation for Economic Co-operation and Development) [26]: "a machine-based system that, for explicit or implicit objectives, infers, from the input it receives, how to generate outputs such as predictions, content, recommendations, or decisions that can influence physical or virtual environments. Different AI systems vary in their levels of autonomy and adaptiveness after deployment". AI system's definition highlighted several distinct aspects such as (1) autonomy, (2) adaptiveness, (3) content generation, (4) decision-making, (5) predictability and (6) recommendation.

AI systems are complex in nature and can be used for

---

This paragraph of the first footnote will contain the date on which you submitted your paper for review, which is populated by XXXX. It is XXXX style to display support information, including sponsor and financial support acknowledgment, here and not in an acknowledgment section at the end of the article.

Asif Q. Gill is with the School of Computer Science, Faculty of Engineering & IT, University of Technology Sydney, Ultimo, NSW 2007, Australia (e-mail: asif.gill@uts.edu.au).

different purposes such as content generation (generative AI or GAI) and decision-making etc. The scope of this paper is limited to AI enabled decision-making, which is a complex human cognitive process. It is worth mentioning that decision-making exists in traditional decision support systems (DSS) as well, however, the new is here the use of the AI capabilities in DSS, such as Machine Learning (ML), Deep Learning (DL), Image Processing (IM) and Natural Language Processing (NLP) to "process large volumes of data, recognize patterns, and generate actionable insights, thus transforming the decision-making landscape across various industrial sectors" [55]. AI enabled decision-making and automation have several economic and societal implications such as they can be used for automating loan approval decisions in financial service sector, optimizing stock and inventory level management decisions in retail sector, optimizing land irrigation and fertilization decisions in agriculture sector, enhancing court decision and sentence consistency in a judicial sector, enhancing driver safety and comfort in automobile sector, and improving quality of patient services and saving lives in health care sector [56]. In a nutshell, (1) the distinct attributes of AI systems, (2) heavy reliance on data and complex processing using ML, DL, IM NLP, and (3) post-deployment unpredictability requires enhancing the agile system development life cycle (SDLC) for supporting AI systems. Thus, this paper draws our attention to the following research question:

RQ: How can agile SDLC be enhanced to support AI systems?

### B. Proposed Solution

This paper uses the decision science [44] and proposes the *human-centric* and *decision-driven* agile SDLC for AI systems to address the above-mentioned research question. Human-centricity means that "human" is the central subject of interest beyond the typical user experience design or shiny AI technology adoption. Based on the decision science, decision-driven refers to the need for understanding "human" decision-making elements and needs rather merely an AI technology-centric push approach. It is important to note here that there are several "AI for agile SDLC" studies, solutions and publications [57-62] that focus on AI-assisted and data-driven software process automation, agility, productivity and quality enhancement etc. However, here the focus is strictly on "agile SDLC for AI systems" and decision-making.

### C. Distinct Contribution

The distinct contribution of this paper is the proposal of the "decision architecture" based on the decision science literature. A decision architecture presents a set of decision elements and artefacts that are critical to analyse and conceptualize the fundamental building blocks of human decision-making such as decision maker, frame, alternatives, logic etc., which can be enabled by AI systems. The proposed decision architecture domain will be discussed within the overall context of human-centric and decision-driven agile SDLC for AI systems. As an indicative proof of concept, the applicability of the proposed agile SDLC for AI systems is demonstrated with the help of an AI enabled decision-making example scenario for insurance claim processing. This initial work lays foundation for further work in this important area of agile SDLC for AI systems and underpinning decision architecture.

In summary, the structure of the paper is as follows. Firstly, it discusses the research background and problem. Secondly, it describes the agile SDLC and underpinning decision architecture for AI systems and demonstrates its application with the help of an insurance claim processing example scenario. Finally, it concludes with key insights and future work options.

## II. RESEARCH BACKGROUND AND PROBLEM

### A. AI System

AI systems are getting significant attention from academia and industry. AI systems attempt to exhibit human-like capabilities such as speech, vision, voice, decision-making and problem solving [27]. There are several examples of AI systems such as AI agents, chatbots, content generation, code generation, digital assistants, decision automation and process automation. AI systems can be applied in various industries such as banking, government, health, higher education and manufacturing for improving user experience and productivity [28-32, 56]. AI systems seem to offer several benefits and there are a range of technological solutions available to choose from (e.g. AWS, Microsoft, Google, IBM). However, there are several challenges, which are attributed to AI systems such as algorithmic and data bias, hallucination, misuse, privacy, safety and trust [33,34]. Though AI assurance, regulations, policies, principles, guidelines and standards are being proposed to address such challenges [35-38], however, there is a lack of full scale SDLC for AI systems, which is important for their trustworthiness.

A typical AI system development is largely data and model-driven, and is composed of stages such as Problem Definition, Data Acquisitions and Preparation, Model Development and Training, Model Evaluation and Refinement, Deployment and Operations [39, 54]. There are ongoing efforts to bring AI system under the DevOps umbrella via practices such as AIOps (AIOps includes related DataOps, MLOps, and ModelOps) [40]. However, these efforts are still very much data, model and technology-centric, and research is required in human-centric full scale agile SDLC for the successful implementation of trustworthy AI systems. This is because there are increasing concerns about the high failure rate of AI projects. For instance, it has been reported that "more than 80 percent of AI projects fail — twice the rate of failure for information technology projects that do not involve AI" [41]. Some of the root causes of such failures are:

- misunderstanding or miscommunication of stakeholders' problem,

- more focus is on using the latest AI technology than understanding and solving the actual problems,

- inability to deploy AI systems,
- lack of appropriate data, and
- apply AI to too difficult problems.

In order to address the above mentioned challenges such as AI bias, hallucination and high failure rates etc., we need a full scale agile SDLC for trustworthy AI systems, and shift our focus to human-centric approaches.

### B. Agile SDLC

There are several agile software development methods (e.g. [3-23]). They have similarities and differences around specific practices (techniques) [16], however, they share same core agile values and principles [2]. They focus on delivering software in short iterations and increments [11,12]. Each iteration or sprint could be of 1-4 weeks duration. An increment or release could be of 1-3 months duration depending on the nature and complexity of a project. A generic or typical agile SDLC can be organised into six key stages [42, 43]: Initiate, Discover, Develop, Operate, Govern, Adapt. An initiative triggers the Initiate stage, which focuses on defining the agile project vision and scope for the software system. Discover stage, which is called here release 0, focuses on iterative planning, analysis, architecture, and design spikes. Design spike refers to an exploratory or investigative activity for designing and prototyping a complex feature with a view to understanding any planning, estimation and development risks, blind spots and solution options. Design spikes may prototype a complex user interface, business logic, and data integration etc. Design spikes are considered useful for early user feedback instead of detailed documentation and big-upfront-design [2].

Develop stage includes DevOps for the iterative software development, testing, continuous integration (CI) and continuous deployment (CD) into operations [14]. Discover and DevOps are connected via feedback loops for adaptation. Operate stage refers to the use of the software system in a live production environment. It also includes system access, support, security, privacy, patching etc. Govern and Adapt stages focus on guiding, monitoring, observing, tracking and handling changes across the agile SDLC. Agile SDLC seems reasonable for traditional software system architecture, design and requirements, which are captured via user stories. Agile SDLC also introduced different variations such as behavior-driven development [4] and test-driven development [13] for enhancing software quality. As discussed earlier, while existing agile SDLC and related approaches work well for the traditional software systems, there is a need to enhance them for AI systems [24], in particular, for decision automation. Here, one may argue "Why can agile SDLC be enhanced to support AI systems". This is because, agile SDLC and related approaches underpin flexible agile values and principles [2] and have advanced to the adaptive enterprise architecture driven large scale agile development [67], which provides a good fit to support the needs of the enterprise scale AI systems and post deployment "adaptability". However, the challenge is how to do so, which is focus of this paper. Here, this paper proposes the use of the decision science to enhance the agile SDLC for AI systems.

### C. Decision Science

Decision science is the study of decision-making. Decisions are all that matter otherwise it is business as usual without any intervention or change. Furthermore, we can only control decision-making quality and not the outcomes. Decision-making can be defined as a complex process, which is about selecting between two or more alternatives or choices that require irrecoverable resources [44]. The core of this paper is to design human-centric and decision-driven agile SDLC for AI systems. Thus, this research draws on the decision theory literature and underpinning approaches [44-46]. Decision theory is concerned with identifying optimal decisions, from available choices under uncertainty, by a rational agent (normative or prescriptive view). Decision theory is also helpful in describing observations about how do humans (people) use information and make decisions (behavior or process) within given constraints or rules (descriptive view)? These prescriptive and descriptive views seem useful for human-centric decision analysis, and decision architecture design. Thus, Agile SDLC can be enhanced by including decision architecture for AI systems.

Four types of decision-making process approaches have been mentioned in decision theory literature [53]: (1) intuitive judgements, (2) rules and shortcuts, (3) importance weighting, and (4) value analysis. The choice of the decision-making process approach depends on decision process goals such as maximize accuracy and transparency or minimize effort and emotional strain. Additionally, four broader types of decisions have been reported in the decision theory literature: (1) choice under uncertainty, (2) cost-benefit, (3) social, and (4) complex decisions. A rational decision maker may arrive at a decision by evaluating the available choices under uncertainty (probabilistic positive or negative outcomes) and cost-benefit analysis [5]. These choices and ultimate final decision could be influenced by human judgements [45] and biases [46]. Judgment is the "human ability to infer, estimate, and predict the character of unknown events" [45]. Human judgements (e.g. prediction of time, cost, quality, expected value or loss), while not perfect, can be augmented via AI systems. Judgment may have hidden biases (e.g. errors in thinking based on previous experience, emotions, intuition heuristics or perception) that need to be carefully identified and treated via debiasing techniques including detailed data analysis.

There are several types of biases such as overconfidence bias, availability bias and confirmation bias [46]. Lot has been discussed about AI bias [33-38]. However, there is also the issue of human bias in decision-making, where AI can be used to debias. We need to look at the duality of the biases both in human decision-making and AI enabled automated decision-making. Thus, it is important to carefully identify, analyse and capture decision-making elements such as (1) human decisions, (2) decision-making process, (3) decision information, (4) judgements, and (5) biases in the early stages of agile SDLC, and ensuring biases are not propagated to AI system development and operations. At the same time, AI systems and decisions need to be thoroughly tested, audited and treated for human as well as algorithmic and data biases. This provides the rational for including the decision architecture, based on the decision science literature, in the agile SDLC for trustworthy AI systems, which will be discussed further in the next section.

## III. AGILE SDLC FOR AI SYSTEMS

This work draws on the literature from decision science [44-46] (e.g. decision architecture) for enhancing agile SDLC [3-24] (e.g. Stages: Initiate, Discover, Develop, Operate, Govern, Adapt) for AI systems [39,40] (e.g. decision-making). Overall, embedded decision architecture is core to the proposed agile SDLC approach for AI systems (Fig. 1). Using the proposed agile SDLC approach, human-centric decision architecture is realized or implemented by the data and model-driven AI systems with a view to achieving decision automation and adaptation. This paper demonstrates the proposed approach application with the help of an insurance claim processing decision-making example scenario.

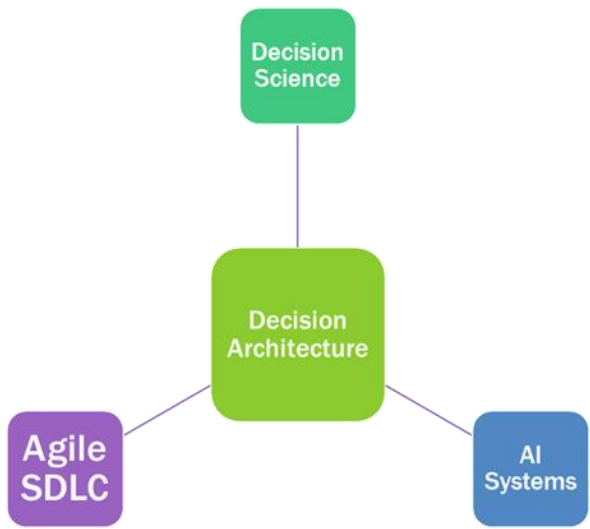

**Fig. 1.** Agile SDLC, decision architecture and AI systems

Agile SDLC for AI systems has two parts (Fig. 2): AI-enabled agile SDLC (bottom-part) and agile SDLC for AI systems (top-part). AI-enabled is where AI copilots can assist humans across the SDLC from initiating an AI system development project to operations including ongoing AI governance, pre- and post-deployment adaptations (a distinct feature of AI systems) [26]. There is a principal-agent relationship between human and AI copilot. However, this is not the focus of the paper. Thus, this section only discusses the top-part agile SDLC for AI systems while having focus on human decision-making. The rational is, as noted from the AI project failure root causes [41], we need to understand human or stakeholder problems before jumping on the technology bandwagon or technology-centric AI adoption. One such area is complex human decision-making and its automation via AI systems. Therefore, a human-centric and decision-driven agile SDLC is proposed for AI systems (Fig. 2). This is organised into six key stages and is explained with the help of an insurance claim processing example.

### A. Insurance Claim Processing Scenario

Before discussing the agile SDLC for AI systems, this section discusses the insurance claim processing scenario (based on [51]). This sets the context and aids in the explainability and applicability of the proposed approach. Insurance claim processing involves the collection and analysis of insurance claim forms and several reports such as incident report, damage evaluation (e.g. water leakage, storm, flood), repair estimates, police report (if relevant), visual images of damages and information about the payouts etc. Insurance claim processing and decision-making require access to the right information at the right time. Manual handling of claim information and decision-making is labor-intensive, expensive, inconsistent and prone to human errors. Further, inability to timely detect fraudulent or valid claims can adversely impact customer experience due to delays in claim processing.

TABLE I
INITIATE: DECISION PROBLEM STATEMENT ELEMENTS

| # | Items | Examples |
|---|-------|----------|
| 1 | Stakeholders | Decision Maker: Claims Manager |
| 2 | Business context | Home insurance, 2 million customers |
| 3 | Business problem | Fraud, delay in claim processing |
| 4 | Business goals | Reduce fraud and operational cost |
| 5 | Business process | Home insurance claim processing |
| 6 | Decision problem | Unclear about which parts (decision-making) of the home insurance claim processing will be automated by using the AI system? |
| 7 | Decision process | Intuitive judgments, rules and shortcuts, important weighting, value analysis |
| 8 | Decision process goals | Maximize decision accuracy and transparency |
| 9 | Budget | $2 million for the project |
| 10 | Time | 6 months |
| 11 | Team | Insurance business and IT |
| 12 | Approach | Decision-driven agile SDLC for AI system development |

### B. Initiate

Agile SDLC stages for AI system can be used for automating decision-making for the insurance claim processing example scenario. Initiate stage focuses on defining the decision problem statement. Drawing on the decision science literature [44-47], initiate stage identified the following decision problem statement elements (project vision and scope) for decision-driven agile AI system development for insurance scenario (Table I). This also addresses the AI project failure root causes, as noted in [41], by understanding the stakeholders' decision problem and is not pushing the data model-driven or latest AI technology adoption agenda. The completion of the initiate stage triggers the discover stage.

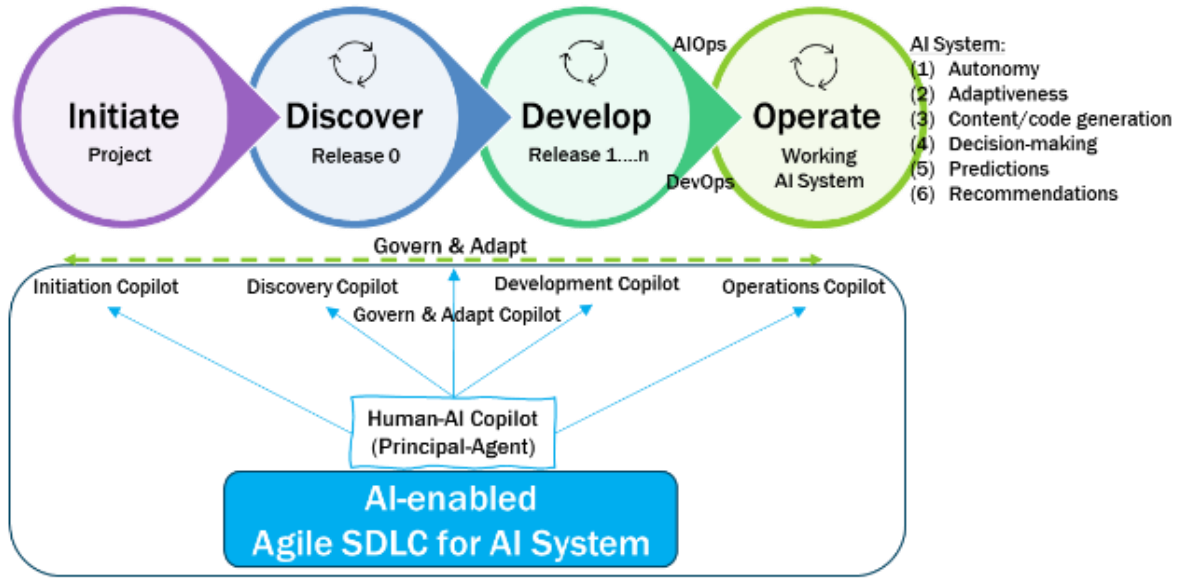

**Fig. 2.** Agile SDLC for AI systems

### C. Discover

Discover stage is organized into five key areas: decision architecture, requirements engineering, AI system architecture, design spikes and plan. All these connected areas of discovery are presented in Fig. 3. This stage further analyses the decision problem and defines the decision architecture independent of any AI technology. Decision architecture describes the decision elements, their relationships to each other and their environment, and principles of decision architecture design and evolution (based on [48]). Decision architecture is important to understand decision-making needs for insurance claims (e.g. decision maker, process, information). This also helps identifying the inherent decision biases (e.g. reject claims lodged from certain locations and demographics, too large a payout), mitigating them by using debasing techniques (e.g. education, intervention, motivation) [50], and mapping how the decision will be made in future by using AI system for insurance claims.

Decision architecture elements are captured as decision requirements; thus, decision architecture and requirements engineering are tightly integrated. Drawing on the decision science, a core set of ten decision elements is identified below [44-46] (Table II), which can be captured in a Decision Catalog [49]. Decision Catalog is a key artifact that captures a list of decisions and underpinning elements that are subject to automation. Additional elements can also be considered, if required, such as decision context, domain, time, cost, criticality, outcomes, limitations and risks. For simplicity reasons, only core decision architecture elements are listed here. Further, these elements can be used to define decision architecture artefacts as noted in Table III. Additional artifacts can be defined depending on the stakeholders' needs and decision complexity and criticality. A decision modelling standard such as DMN (Decision Model and Notation) can also be used to model the decision architecture artifacts [64]. Agile teams may use simple arrows and boxes or what-if-analysis or computer software generated simulations for modelling the relevant decision architecture artifacts or diagrams. This paper does not restrict the use of formal or informal modelling and notations for decision architecture.

TABLE II
DECISION ARCHITECTURE CORE ELEMENTS

| # | Items | Examples |
|---|-------|----------|
| 1 | Decision maker | Claims Manager |
| 2 | Frame | Decision maker's problem and goal |
| 3 | Alternatives | Choices or options to choose from for when processing insurance claims |
| 4 | Preferences | Anchors or priorities when processing insurance claims |
| 5 | Information | Information available for decision-making when processing insurance claims |
| 6 | Decision logic | Decision process, algorithms, calculations, models, reasoning |

| 7 | Decision rule | Business rules and knowledge |
| 8 | Bias | Human and algorithm biases, systematic errors and short cuts when processing insurance claims and making decision |
| 9 | Principles | Fair, consistent, nonmaleficence decisions |
| 10 | Automation Level | Manual, semi-automated, automated, autonomous insurance claim processing |

TABLE III

DECISION ARCHITECTURE ARTIFACTS

| # | Items | Examples |
|---|-------|----------|
| 1 | Decision Org Chart | Capture decision-making roles, accountabilities and responsibilities e.g. endorser, maker, follower |
| 2 | Decision Canvas | Capture specific elements such as decision problem, process goals, judgements, biases, debiasing techniques, choices |
| 3 | Decision Card | Capture each decision requirements without detailing decision logic or how a decision will be made |
| 4 | Decision Prompt | Capture the details that will be provided to the AI systems during interactions with AI system |
| 5 | Decision Catalog | List of decisions and underpinning elements. |
| 6 | Decision Hierarchy | Capture strategic, tactical, operational level decisions |
| 7 | Decision Process Model | Decision-making process or workflow |
| 8 | Decision Service | A reusable set of decisions and related activities that can be invoked by the decision process |
| 9 | Information Model | Information elements and their relationships as input for decision making |
| 10 | Rule Model | Decision rules and their relationship for supporting decision making |
| 11 | Enterprise Knowledge Graph | Business knowledge for supporting decision making and capturing decision context, reasoning, explanation |
| 12 | Decision Table | Capture decision logic in a tabular format |
| 13 | Decision Matrix | Capture decision logic in a matrix format |
| 14 | Decision Tree | Capture decision logic in a tree format |

Decision architecture elements and related artifacts provide a solid foundation and initial requirements for informing the human-centric and bias free trustworthy AI system architecture, design spikes and overall agile planning for AI system development in short releases and underpinning iterations. Requirements engineering [23] is a central activity that provides a systematic process for the iterative elicitations, analysis, specification and management of decision requirements in the requirements backlog. Each decision requirements can be written on a decision card [65] and stored in the decision requirements backlog or repository (Table III). Each decision card can capture details such as (1) what a decision is about, (2) why a decision needs to be taken, (3) who needs to be involved, (4) who will be impacted, (5) when a decision needs to be taken, and (6) what is the required automation level? Each decision card is prioritized, estimated, planned and used for designing the AI system architecture and agile planning. Related decision cards can be grouped where there is a dependency (e.g. decision hierarchy or clustering) between decision requirements. Each decision card represents one decision. Here, we can also design decision prompts that will be used for interacting with AI systems and provide the decision questions and related instruction, examples, information, knowledge, and context etc. These prompts can be designed using different prompts frameworks such as RTF (role, task, format), Chain of Thoughts, Chain of Density, RISEN (role, instructions, steps, end goal, narrowing), and RODES (role, objective, details, examples, sense check) [69]. Agile plan is organized into release, where each release can have several iterations. Estimated and prioritized decision cards are assigned to different releases and iterations for the iterative and incremental development of AI systems.

AI system architecture consists of integrated application, data and algorithm architectures [68]. It provides the initial design of the AI-enabled automated decision-making for insurance claim processing (not big upfront design). AI system contains claim processing software applications and underpinning software agents [52] (e.g. application architecture) that interact with each other and their environment to collect and process insurance claim data (e.g. data architecture) for decision making using the AI algorithms (e.g. algorithm architecture). AI system architecture (including decision architecture) evolved as the development proceeds in short releases. Design spikes can be used to develop prototypes for AI system requirements clarification and risk identification. To support decision requirements (noted on decision cards), AI system requirements, as a results of AI system architecture and design spikes, are captured in the requirements backlog via the requirements engineering process. As an example, for a fully automated decision-making scenario, AI system architecture flow is given below - architecture elements are highlighted as well (based on [51]):

- *Users* are prompted to upload the insurance claim forms and reports using the insurance *application* via their *interface (e.g. prompt)*.

- Insurance *data* is automatically extracted from the forms and loaded into the data system.

- Software *agent (data processor)* is activated that engages the *AI algorithm* to process the captured claim data.

- *AI algorithm* uses the claim data as an input and creates data summary and key data points for decision making.

- Software *agent (decision generator)* based on the output of the AI algorithm, automatically generates a decision to accept or reject the claim or ask for more information without human intervention.

- Software agent (*decision coordinator*) coordinates other agents' activities.

- While software agent has assisted the decision maker (*human*) with the proposed generated decision, however, final endorsement and responsibility still sits with the human decision maker.

The details of the AI system architecture can be designed using trimodal thinking [68] and visualized using the C4 Model approach [70]. Trimodal thinking helps to organize the AI system architecture components based on the (1) intuitive, (2) rational and (3) control thinking for decision-making. For instance, intuitive components of the AI system use patterns-based decision making and related AI algorithms and models such as Large Language Models (LLMs) for decision automation and prompting. Data-driven complex calculations and rules-based decision automation use the rational components of an AI system. Control provides the required governance, quality, observability and monitoring of intuitive and rational components of an AI system. C4 Model based approach can be used to detail the inner details of the AI system architecture stack at four levels - from high to low detailed levels: (1) AI system (context), (2) container, (3) component and (4) code (Fig. 4). Based on C4 model [70], an AI system is composed of one or more containers (applications, data & algorithms). Each container has one or more components. For instance, an application is a container that can have several agents and services. Each component can be implemented by one or more code elements such classes, objects and functions etc. Actors or users use the AI systems. Depending on the complexity and stakeholders' needs, an AI system architecture can be designed at a very high contextual level or low detailed level.

Given the focus of this paper is on decision architecture, thus, additional details about the AI system architecture can be discussed in future articles. It is important to mention that AI assurance, regulations, policies, principles, guidelines and standards can be used to assess the quality of the proposed decision architecture and AI system architecture for quality assurance and compliance purposes [35-38]. This is typically achieved via the architecture review board [43]. An initial agile plan is developed, where decision automation requirements (e.g. based on decision-making and AI systems requirements) are allocated to different iterations and releases as a starting point. This plan evolves as the development stage is executed and can be modified adjusted as required (typical agile practice). In summary, discover stage is also referred as release 0, and provides a good foundation for the next stage. It is important to mention here that instead of a big-up-front detailed design, discover stages focus on the minimum required architecture (e.g. connected decision and AI system architecture), which evolves as the AI system development proceeds in short increments.

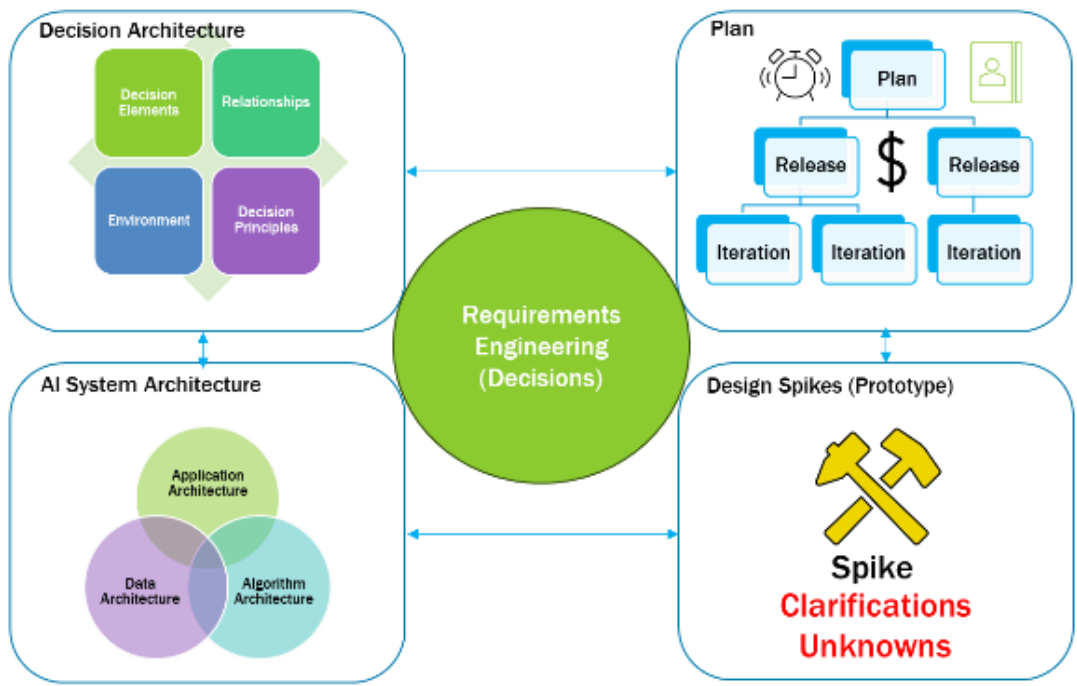

**Fig. 3.** Discover

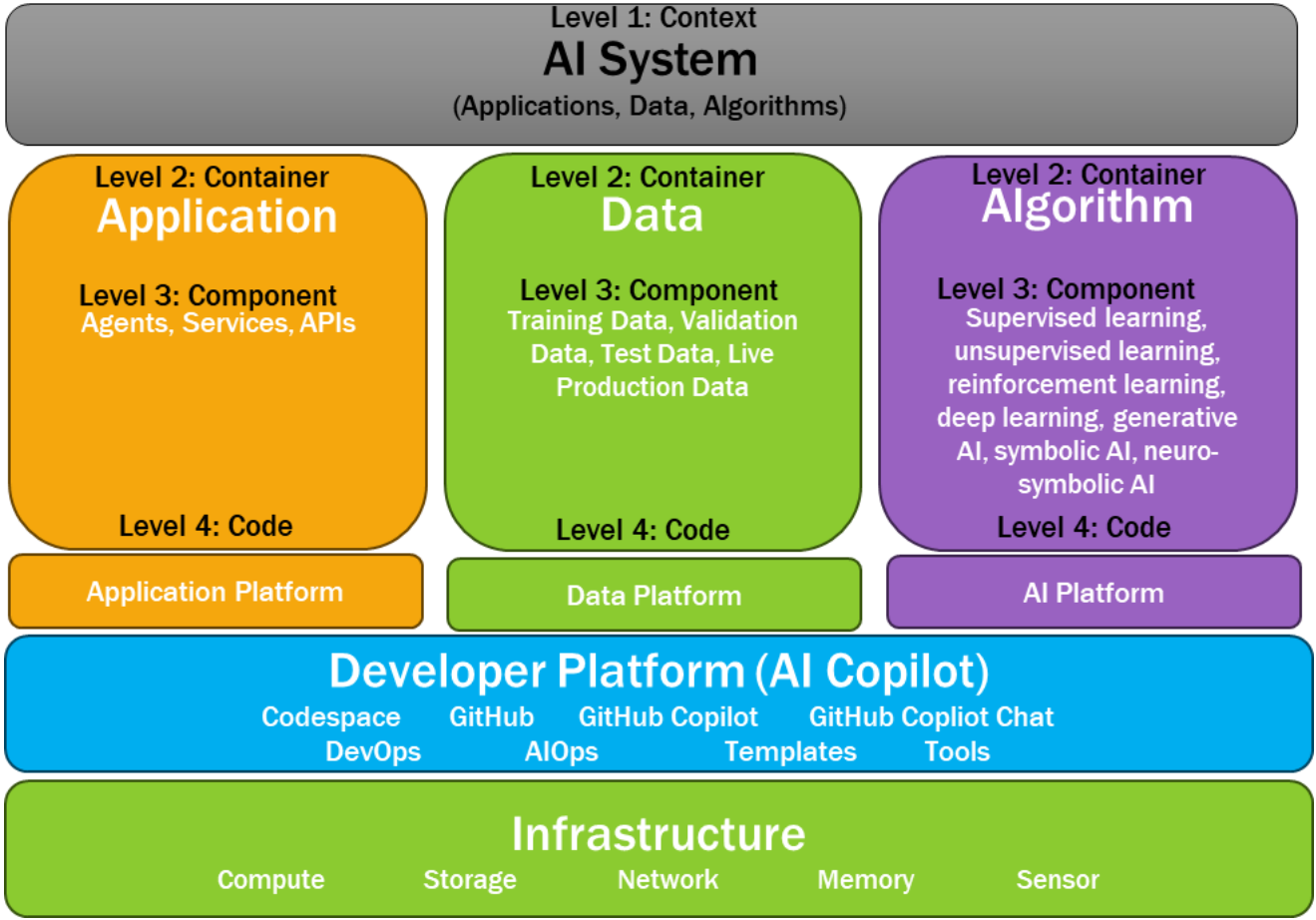

**Fig. 4.** AI system architecture stack and developer platform example

*D. Develop*

Develop stage uses the developer platform (Fig. 4) and iteratively implements the decision architecture and supporting AI system architecture requirements (applications, data and algorithms), and releases the working AI system in operations. This stage involves developing software application components such as agents, services, APIs, and AI models using selected algorithms and data for automated decision-making for insurance claims. For instance, an agent component can be implemented by agent classes using Python programming language and Semantic Kernel open-source development kit. Data could be training, validation, test and live production data. AI algorithms and related models are trained and tested for performance, accuracy and biases in data and AI model. All these components are continuously integrated (CI) and deployed (CD) into operations for end users. Further, AI system requires application, data and AI platforms to host and run the relevant components. Developer platform provides support for well-established DevOps [14] and AIOps practices [39,40], templates and tools (e.g. extensions, functions) [63] It also includes codespace (AI system project), GitHub (code repository and collaboration), GitHub Copilot and Chat for AI assisted AI system development, debugging and testing. This should also include testing of the AI system decision-making behavior, patterns, prompt injections, algorithm and data poisoning, inversion as well as penetration testing algorithms and models [66].

*E. Operate*

Operate stage manages the live working AI system and its use for automated decision-making during insurance claim processing. This involves AI system access management, availability management and capacity management including the handling of incidents, monitoring, observability, patching, privacy, problem, reliability, resiliency, security and trust. For instance, AI systems and decisions should be observable, thus decision log, traces, events and metrics need to be maintained for AI system audit and assurance. Post deployment AI system adaptations, due to changing live production data, must be addressed by the operations as well. This is important because, unlike traditional systems, AI system algorithms grow on data and their accuracy and speed may degrade overtime. Continuous AI assurance and performance assessments are required for smooth operations of the working AI system.

*F. Govern and Adapt*

AI systems grow over data and adapt pre and post deployment. These adaptation needs are governed and responded to in a systematic manner for responsible, safe and trustworthy AI systems [35-38]. Govern stage provides the necessary guidance via architecture review board and different frameworks such as AI system assurance framework, principles etc. Adapt stage focuses on continuous monitoring, tracking and observability over the whole human-centric and decision-driven AI system across the entire agile SDLC. Adapt stage is critical for handling unpredictable situations and decision-making needs.

## IV. DISCUSSION AND CONCLUSION

There is a significant interest in AI systems for decision automation. While AI system adoption seems beneficial, it also poses several challenges such as algorithmic and data bias, hallucination, misuse, privacy, safety and trust. As noted earlier, more than 80% of AI projects fail, which is twice the failure rate of traditional non-AI systems. AI systems have distinct attributes such as autonomy, content generation and decision automation compared to traditional non-AI systems. AI systems are largely data and model driven and demand a full scale SDLC. Agile SDLC seems appropriate for AI projects, however, it needs to be enhanced to support AI system development.

This paper addressed this important need. Based on the decision science literature, this paper proposed a full scale agile SDLC for the AI systems. While agile SDLC has initiate, discover, develop, operate, govern and adapt stages that can be used for AI systems, however, the distinct contribution of this paper is the proposal of the decision architecture domain. Decision architecture is embedded in the discover stage of the proposed agile SDLC for AI systems. Before specifying the decision architecture, decision problem is also defined in the initiate stage of the proposed agile SDLC for AI systems. This is important because decision problem definition and decision architecture are core to the human-centric and decision-driven AI systems. Decision architecture elements and related artifacts have been proposed based on the review of related concepts from decision science literature. This paper discussed the core elements of a decision architecture such as the decision maker, frame, alternatives, preferences, logic, information, bias and automation level etc. These core elements and their relationships can help designing the current and future state of a decision architecture for a given business area or domain such as the insurance claim process within the customer support business area.

Gaps between the current and future state can be captured as decision requirements, which can be managed via the existing agile requirements engineering process. There is also an influx of data and AI regulations and assurance frameworks [35-38]. These compliance and assurance requirements can be baked into the design of the decision architecture. Further, this research has major implications for the future of trustworthy AI systems, solutions and their impact on society. For instance, it draws our attention to identify and address bias using debiasing techniques for addressing any potential unethical issues that may harm humans in the design of the decision architecture. Overall, this paper aims to provide a bias free and technology agnostic view of the human-centric decision architecture. Here, focus is on supporting "humans" and society at large and identifying and describing decisions including relevant decision processes and information that are subject to automation via AI systems. Hence, human-centric decision architecture is a pre-requisite for the trustworthy AI systems architecture and their development.

This paper provides initial insights and needs for the design of a decision architecture, which has not been discussed before. The application and explanation of the proposed approach is offered via an insurance claim processing example scenario. It is anticipated that the proposed decision architecture is useful for enhancing the agile SDLC for AI systems aiming for decision-automation. Learnings from this work can also be applied to enhance relevant enterprise architecture frameworks, which do not seem to have decision architecture as an explicit domain [43]. Enterprise architecture is critical for developing and deploying enterprise scale trustworthy AI systems. This work provides an important initial foundation for further research in this new area of decision architecture and agile SDLC for AI systems. Research can be advanced to further refine the decision architecture elements and artifacts. New elements and artifacts can be discovered or existing can be modified as per the specific context. Similarly, future research may investigate into the development of domain specific decision architecture ontology, architecture patterns and supporting solutions.

## ACKNOWLEDGMENT

I would like to extend sincere thanks to colleagues from academia and industry for providing their valuable feedback in improving the quality of this paper.

## REFERENCES

[1] E.F. Casali, "A brief history of agile methods". Intense Minimalism, 2012. Available: https://intenseminimalism.com/2012/a-brief-history-of-agile-methods/

[2] Agile Alliance, Manifesto for Agile Software Development, 2001. https://agilemanifesto.org/.

[3] J.A. Highsmith, Adaptive Software Development: A Collaborative Approach to Managing Complex Systems, New York: Dorset House, 2002.

[4] J.F. Smart, BDD in Action: Behavior-Driven Development for the Whole Software Lifecycle. Manning Publications, 2014.

[5] A. Cockburn,. Agile Software Development. Addison-Wesley, Boston, 2002.

[6] J. Stapleton, DSDM: The Method in Practice. Addison-Wesley, 1997.

[7] S. Ambler, and M. Lines, Choose Your WoW! A Disciplined Agile Delivery Handbook for Optimizing Your Way of Working, 2019.

[8] K. Beck, Extreme Programming Explained: Embrace Change. Addison-Wesley, ISBN 978-0-321-27865-4, 1999.

[9] S.R. Palmer and J.M. Felsing, A Practical Guide to Feature-Driven Development. Prentice-Hall Inc, Upper Saddle River, 2002.

[10] M. Poppendieck and T. Poppendiec, Lean Software Development: An Agile Toolkit. Addison-Wesley Professional, 2003.

[11] K. Schwaber, and M. Beedle, M. Agile Software Development with SCRUM. Prentice Hall, 2002.

[12] D. Leffingwell, Scaling Software Agility: Best Practices for Large Enterprises. Addison-Wesley, 2007

[13] J.W. Newkirk, and A.A. Vorontsov, Test-Driven Development in Microsoft .NET, Microsoft Press, 2004.

[14] L. Bass, I. Weber, and L. Zhu, DevOps: A Software Architect's Perspective. Addison-Wesley, 2015.

[15] P. Abrahamsson, O. SaloJussi, J. Ronkainen, J. Warsta, Agile Software Development Methods: Review and Analysis, VTT Technical Research Centre of Finland, VTT Publications 478, Otamedia, 2002. Available: https://arxiv.org/abs/1709.08439

[16] A. Qumer, and B. Henderson-Sellers, "An evaluation of the degree of agility in six agile methods and its applicability for method engineering.", Information and Software Technology, vol. 50, no. 4, 2008.

[17] A. Qumer, and B. Henderson-Sellers, "A framework to support the evaluation, adoption and improvement of agile methods in practice.", Journal of Systems and Software vol. 81, no. 11, 2008.

[18] Y.I. Alzoubi, A.Q. Gill, and A. Al-Ani, "Empirical studies of geographically distributed agile development communication challenges: A systematic review." Information & Management, vol. 53, no. 1, 2016.

[19] G.B. Ghantous, and A. Gill, "DevOps: Concepts, practices, tools, benefits and challenges.", PACIS2017, 2017.

[20] R. Hoda, N. Salleh, J. Grundy, and H.M. Tee, "Systematic literature reviews in agile software development: A tertiary study.", Information and software technology vol. 85, 2017.

[21] A.Q. Gill, B. Henderson-Sellers, and M. Niazi, "Scaling for agility: A reference model for hybrid traditional-agile software development methodologies.", Information Systems Frontiers, vol. 20 2018.

[22] L. Leite, C. Rocha, F. Kon, D. Milojicic, and P. Meirelles, "A survey of DevOps concepts and challenges." ACM Computing Surveys (CSUR), vol. 52, no. 6, 2019.

[23] J. Buchan, D. Zowghi, and M. Bano, "Applying Distributed Cognition Theory to Agile Requirements Engineering.", In Requirements Engineering: Foundation for Software Quality: 26th International Working Conference, REFSQ 2020, Pisa, Italy, March 24–27, 2020.

[24] D. Zowghi and M. Bano, "What's Missing in Requirements Engineering for Responsible AI?," in IEEE Software, vol. 40, no. 6, pp. 11-15, Nov.-Dec. 2023.

[25] J.L. McCarthy, M.L. Minsky, N. Rochester, and C.E. Shannon, "A Proposal for the Dartmouth Summer Research Project on Artificial Intelligence", 1956.

[26] M. Grobelnik, K. Perset, and S. Russell, What is AI? Can you make a clear distinction between AI and non-AI systems? OECD, 2024.

[27] Z. Tekic, and J. Füller, "Managing innovation in the era of AI," Technology in Society, vol73, 2023, https://doi.org/10.1016/j.techsoc.2023.102254.

[28] O. Sospeter, M. Finke, J. Belke, F. Dyck, and C. Kürpick., "Use Case Catalog and Assessment for AI Applications in Intralogistics of Manufacturing Companies." Procedia CIRP, vol. 118, pp. 74-79, 2023.

[29] K. Michael, J. Pitt, J. Sargent and E. Scornavacca, "Automating Higher Education Through Artificial Intelligence?," in IEEE Transactions on Technology and Society, vol. 5, no. 3, pp. 264-271, Sept. 2024,

[30] M. Chui, J. Manyika, M. Miremadi, N. Henke, R. Chung, P. Nel, and S. Malhotra, "Notes from the AI frontier: Insights from hundreds of use cases." McKinsey Global Institute 2, 2018.

[31] M. Tarafdar, C.M. Beath, and J.W. Ross, "Using AI to enhance business operations." MIT Sloan Management Review 60, no. 4, 2019.

[32] K. Alhosani, and S.M. Alhashmi, "Opportunities, challenges, and benefits of AI innovation in government services: a review." Discover Artificial Intelligence 4, no. 1, 2024.

[33] K. Michael, J. R. Schoenherr and K. M. Vogel, "Failures in the Loop: Human Leadership in AI-Based Decision-Making," in IEEE Transactions on Technology and Society, vol. 5, no. 1, pp. 2-13, March, 2024, doi: 10.1109/TTS.2024.3378587.

[34] I.H. Sarker, H. Janicke, A. Mohsin, A. Gill, and L. Maglaras, "Explainable AI for cybersecurity automation, intelligence and trustworthiness in digital twin: Methods, taxonomy, challenges and prospects." ICT Express, 2024.

[35] EU. The EU Artificial Intelligence Act. Available: https://artificialintelligenceact.eu/

[36] Australian Government. Policy for responsible use of AI in government. Available: https://architecture.digital.gov.au/responsible-use-of-AI-in-government

[37] OECD. AI Principles overview. Available: https://oecd.ai/en/ai-principles

[38] NIST. AI Standards. Available: https://www.nist.gov/artificial-intelligence/ai-standards

[39] J. Saltz, What is the AI Life Cycle?, Data Science Process Alliance, 2024. Available: https://www.datascience-pm.com/ai-lifecycle

[40] S. Hegde, All the Ops: DevOps, DataOps, MLOps, and AIOps, IBM, 2023. Available: https://developer.ibm.com/articles/all-the-ops-devops-dataops-mlops-and-aiops/

[41] J. Ryseff, B.F. De Bruhl, and S.J. Newberyy, "The Root Causes of Failure for Artificial Intelligence Projects and How They Can Succeed: Avoiding the Anti-Patterns of AI", RAND, 2024.

[42] A.Q. Gill, and E. Chew, "Configuration information system architecture: Insights from applied action design research." Information & Management 56, no. 4, 2019.

[43] A.Q. Gill, "Adaptive enterprise architecture as information: Architecting intelligent enterprises". World Scientific Publishing, Singapore, 2022

[44] R.A. Howard, A.E. Abbas, "Foundations of Decision Analysis". Pearson, 2016.

[45] R. Hastie and R.M. Dawes. Rational choice in an uncertain world: the psychology of judgment and decision making. California: Sage Publications, 2001.

[46] M. G. Haselton, D. Nettle and P.W. Andrews, "The evolution of cognitive bias". In The handbook of evolutionary psychology. Buss, D.M., Ed. New Jersey: Wiley Online Library. 968-987, 2015.

[47] B. Schwartz and A. Ward, Maximizing versus satisficing: Happiness is a matter of choice, 2002.

[48] ISO/IEC/IEEE 42010. Defining architecture. Available: http://www.iso-architecture.org/ieee-1471/defining-architecture.html

[49] A.Q. Gill and M. Hansnata, "Digital Government Ecosystem: Adaptive Architecture for Digital and ICT Investment Decision Making.", In Proceedings of the 25th Annual International Conference on Digital Government Research, pp. 555-564. 2024.

[50] J.B. Soll, K.L. Milkman, and J.W. Payne, "A user's guide to debiasing. In The Wiley Blackwell handbook of judgment and decision making", vol. II. Oxford: Blackwell Publishing. 2015.

[51] Expert.ai. Claims Automation. Available: https://www.expert.ai/products/claims-automation/

[52] A. Qumer and B. Henderson-Sellers, "A framework to support non-fragile agile agent-oriented software development,", SoMeT (2006): 84-100.

[53] P. Schoemaker, and J.E. Russo, A pyramid of decision approaches. California Management Review, vol. 36, 9-31, 1993.

[54] S. Amershi S., A. Begel, C. Bird, R. DeLine, H. Gall E. Kamar, N. Nagappan, B. Nushi B, and T. Zimmermann, "Software engineering for machine learning: A case study", In 2019 IEEE/ACM 41st International Conference on Software Engineering: Software Engineering in Practice (ICSE-SEIP), 291-300, 2019.

[55] M. Soori, F.K.G. Jough, R. Dastres, and B Arezoo, "AI-Based Decision Support Systems in Industry 4.0, A Review," Journal of Economy and Technology, 2024.

[56] Intellias. AI Decision Making: What Is It, Benefits & Examples, 2025. Available: https://intellias.com/ai-decision-making/.

[57] M. Coutinho, L. Marques, A. Santos, M. Dahia, C. Franca, and R.S. Santos, "The Role of Generative AI in Software Development Productivity: A Pilot Case Study.", arXiv preprint arXiv:2406.00560, 2024.

[58] D. Glushkova, "The influence of Artificial intelligence on productivity in Software development." PhD diss., Politecnico di Torino, 2023.

[59] M.A. Hassan, "Impact of adopting AI tools by software developers towards productivity and sustainability.", 2024.

[60] A,M. Dincă, A.M., SD. Axinte, G. Tod-Raileanu, and I.C. Bacivarov, "AI Tools introduced in Software Development. Analysis of Code quality, Security and Productivity Implications", In 2024 IEEE 30th International Symposium for Design and Technology in Electronic Packaging (SIITME). 32-39, 2024.

[61] D. Ajiga, P.A. Okeleke, S.O. Folorunsho, and C. Ezeigweneme, "Enhancing software development practices with AI insights in high-tech companies." IEEE Software Engineering Institute, Technical Report TR-2024-003, 2024

[62] J. Sauvola, S. Tarkoma, M. Klemettinen, J. Riekki, and D. Doermann, "Future of software development with generative AI." Automated Software Engineering 31, no. 1, 2024.

[63] J. Wiesinger, P. Marlow, and V. Vuskovic. Agents, Google, 2024. Available: https://medium.com/@penkow/summary-of-googles-ai-white-paper-agents-d5670ae495c9

[64] OMG Stadnards Development Organization, Decision Model and Notation Version 1.6 Beta 1, 2024. Available: https://www.omg.org/spec/DMN

[65] M. Gutierrez Lopez, G. Rovelo Ruiz, K. Luyten, M. Haesen, and K. Coninx, "Re-thinking Traceability: A prototype to record and revisit the evolution of design artefacts." In Proceedings of the 2018 ACM International Conference on Supporting Group Work, 196-208. 2018.

[66] J. Brownlow Davies, Introducing AI Penetration Testing, Bugcrowd, 2024. Available: https://www.bugcrowd.com/blog/introducing-ai-penetration-testing

[67] A.Q. Gill, "Adaptive enterprise architecture drivenagiledevelopment." In 2015 International Conference on Information Systems Development, 2015.

[68] A. Gill, "Trimodal Thinking for Architecting Human-Centric AI Systems: Fast, Slow and Control." Authorea Preprints, 2024.

[69] M. Kremb, 5 prompt frameworks to level up your prompts: RTF, RISEN, RODES, Chain of thought and Chain of density, 2023. Available: https://www.thepromptwarrior.com/p/5-prompt-frameworks-level-prompts

[70] C4 Model, The C4 model for visualising software architecture. Available: https://c4model.com/

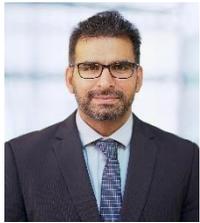 **Asif Q. Gill** (Senior Member, IEEE) is Professor and Head of Discipline Software Engineering at the School of Computer Science, University of Technology Sydney. He is also the Director of the DigiSAS Lab. He has a PhD in Computing and MSc Computing Science. He is a member of the ACS Data Sharing Committee, IFIP Technical Committee 8.1, and Standards Australia Software and Systems Engineering Committee IT-015. He is also an associate editor of the IEEE Transactions on Technology & Society and Springer Nature Discover Data journals. He is often invited and involved as a professional keynote speaker, editor, conference chair, organizer and reviewer for several national and international academic and industry conferences.

He authored 3 books and 180+ articles. His work has appeared in major academic journals: IEEE Transactions on PC, Information and Management, Information Systems, CAIS, Computers and Security, Data and Knowledge Engineering, Information and Software Technology, Information Systems Frontiers, Journal of Systems and Software.